\documentclass[aip,amsmath,amssymb,reprint]{revtex4-2}

\usepackage[final]{graphicx}
\usepackage{float}
\usepackage{hyperref}
\hypersetup{colorlinks=true, citecolor=blue, urlcolor=blue, linkcolor=blue}
\usepackage{color}	
\newcommand{\tred}[1]{\textcolor{black}{#1}}

\makeatletter
\def\@email#1#2{%
 \endgroup
 \patchcmd{\titleblock@produce}
  {\frontmatter@RRAPformat}
  {\frontmatter@RRAPformat{\produce@RRAP{*#1\href{mailto:#2}{#2}}}\frontmatter@RRAPformat}
  {}{}
}%
\makeatother

\begin{document}

\author{Nicolas Gauthier}
\email{nicolas.gauthier@inrs.ca}
\affiliation{Institut National de la Recherche Scientifique - Énergie Matériaux Télécommunications Varennes QC J3X 1P7 Canada}

\author{Hadas Soifer}
\affiliation{School of Physics and Astronomy, Tel-Aviv University, Tel-Aviv 69978, Israel}

\author{Jonathan A. Sobota}% ask about position
\affiliation{Stanford Institute for Materials and Energy Sciences, SLAC National Accelerator Laboratory, Menlo Park, California 94025, USA}

\author{Heike Pfau}
\affiliation{Department of Physics, The Pennsylvania State University, University Park, PA 16802, USA}

\author{Edbert J. Sie}
\affiliation{Stanford Institute for Materials and Energy Sciences, SLAC National Accelerator Laboratory, Menlo Park, California 94025, USA}
\affiliation{Geballe Laboratory for Advanced Materials, Departments of Applied Physics and Physics, Stanford University, Stanford, California 94305, USA}

\author{Aaron M. Lindenberg}
\affiliation{Stanford Institute for Materials and Energy Sciences, SLAC National Accelerator Laboratory, Menlo Park, California 94025, USA}
\affiliation{Department of Materials Science and Engineering, Stanford University, Stanford, CA 94305, USA}
\affiliation{Stanford PULSE Institute, SLAC National Accelerator Laboratory, Menlo Park, CA 94025, USA}

\author{Zhi-Xun Shen}
\affiliation{Stanford Institute for Materials and Energy Sciences, SLAC National Accelerator Laboratory, Menlo Park, California 94025, USA}
\affiliation{Geballe Laboratory for Advanced Materials, Departments of Applied Physics and Physics, Stanford University, Stanford, California 94305, USA}
\affiliation{Department of Physics, Stanford University, Stanford, California 94305, USA}

\author{Patrick S. Kirchmann}
\email{kirchman@stanford.edu}
\affiliation{Stanford Institute for Materials and Energy Sciences, SLAC National Accelerator Laboratory, Menlo Park, California 94025, USA}

\title{Analysis methodology of coherent oscillations in time- and angle-resolved photoemission spectroscopy}

\date{\today}

%%%%%%%%%%%%%%%%%%%%%%%%%%%%%%%%%%%%%%%%%%%%%%%%%%%%%%%%
%%%%%%%%%%%%%%%%%%%%%%%%%%%%%%%%%%%%%%%%%%%%%%%%%%%%%%%%
\begin{abstract}
Oscillatory signals from coherently excited phonons are regularly observed in ultrafast pump-probe experiments on condensed matter samples. 
Electron-phonon coupling implies that coherent phonons also modulate the electronic band structure. 
These oscillations can be probed with energy and momentum resolution using time- and angle-resolved photoemission spectroscopy (trARPES) which reveals the orbital dependence of the electron-phonon coupling for a specific phonon mode. 
However, a comprehensive analysis remains challenging when multiple coherent phonon modes couple to multiple electronic bands. 
Complex spectral line shapes due to strong correlations in quantum materials add to this challenge. 
In this work, we examine how the frequency domain representation of trARPES data facilitates a quantitative analysis of coherent oscillations of the electronic bands. 
We investigate the frequency domain representation of the photoemission intensity and \tred{the first moment of the energy distribution curves}. 
Both quantities provide complimentary information and are able to distinguish oscillations of binding energy, linewidth and intensity.
We analyze a representative trARPES dataset of the transition metal dichalcogenide WTe$_2$ and construct composite spectra which intuitively illustrate how much each electronic band is affected by a specific phonon mode. 
We also show how a linearly chirped probe pulse can generate extrinsic artifacts that are distinct from the intrinsic coherent phonon signal.   
\end{abstract}
\maketitle

%%%%%%%%%%%%%%%%%%%%%%%%%%%%%%%%%%%%%%%%%%%%%%%%%%%%%%%%
%%%%%%%%%%%%%%%%%%%%%%%%%%%%%%%%%%%%%%%%%%%%%%%%%%%%%%%%
\section{Introduction}

Electron-phonon coupling is a fundamental concept in the description of solid state materials and plays a dominant role in quantum phenomena such as superconductivity \cite{Bardeen1957}. 
Femtosecond light pulses can excite phase-locked phonon modes, leading to coherent oscillations of the lattice structure with a frequency corresponding to the phonon energy \cite{Zeiger1992}. 
The coherent structural response can be probed directly with ultrafast diffraction techniques such time-resolved x-ray diffraction \cite{Fritz2007,Gerber2015,Rettig2015,Gerber2017,Trigo2019,Huang2023,Lindenberg2017} or ultrafast electron diffraction \cite{Bugayev2011,Chatelain2014,Waldecker2017,Sie2019,Ungeheuer2024}.
Electron-phonon coupling implies that coherent structural dynamics cause coherent modulations of the electronic bands. 
Time- and angle-resolved photoemission spectroscopy (trARPES) probes these oscillations with energy and momentum resolution \cite{Boschini2024}. 
This orbital resolution in trARPES is crucial for the study of multi-band systems \cite{Sobota2014,Hein2020,Lee2023}.
Recent advances in multi-modal ultrafast pump-probe experiments made it possible to quantify the electron-phonon coupling for specific phonon modes and well-defined electronic bands \cite{Rettig2015,Gerber2017,Sobota2023,Huang2023}. 
Coherent responses in trARPES have been studied for a variety of materials such as metals \cite{Loukakos2007,Rettig2012}, semimetals\cite{Papalazarou2012,Faure2013,Sakamoto2022}, topological insulators \cite{Sobota2014,Golias2016,Sobota2023,Huang2023} 
, transition metal chalcogenides \cite{Perfetti2006,Hellmann2012,Tang2020,Suzuki2021,Maklar2023,Ren2023,Baldini2023}, rare-earth tritellurides \cite{Schmitt2008,Leuenberger2015,Maklar2021}, and unconventional superconductors \cite{Avigo2013,Yang2014,Yang2015,Gerber2017,Yang2019}.

A quantitative analysis of coherent oscillations in trARPES typically consists of three components: oscillations of spectral weight, linewidth or binding energy.
Binding energy oscillations are the basis for quantitative estimates of the deformation potential \cite{Rettig2015,Gerber2017,Huang2023}.
Coherent oscillations in trARPES are commonly analyzed quantitatively by fitting momentum or energy distribution curves (EDCs) as a function of pump-probe delay. 
Fitting is reliable for spectra with few and sufficiently separated bands with well-defined line shapes \cite{Yang2019,Sakamoto2022}. 
However, more complicated line shapes near the Fermi level  and complex spectra in multiband systems \cite{Yang2014,Yang2015,Gerber2017,Hein2020} can limit the fidelity of fits. 
Uncertainties increase due to cross-correlated fit parameters which may introduce systematic errors and conceal weak coherent responses. 
The complexity of trARPES spectra is increasing as the momentum access is expanding with high-harmonic-generation light sources \cite{NA2023,Boschini2024} and bias voltage techniques \cite{Gauthier2021a}. 
Momentum microscopes \cite{Tusche2015,Maklar2021,Karni2022} and related developments \cite{Majchrzak2024} have the potential to record four-dimensional trARPES data sets for which fits may involve a prohibitive effort.
All this suggest that the trARPES community would benefit from a generally applicable, numerical approach to characterize coherent oscillations quantitatively and comprehensively with only minimal adjustments \cite{Xian2023}. 

First steps in this direction have been taken recently  \cite{Giovannini2020,DeGiovannini2022,Hein2020,Suzuki2021,Lee2023,Ren2023,Emeis2024} by presenting a specific frequency through a Fourier transform of the spectral intensity along the time delay axis which is evaluated for each energy and momentum point. 
This approach has been labeled frequency-domain ARPES (FDARPES) and highlights the oscillating features of a band structure at a given frequency.
The FDARPES representation is qualitatively different than conventional trARPES and not straightforward to interpret. 
Theoretical work \cite{Giovannini2020,DeGiovannini2022} demonstrated how features in the frequency domain can be quantitatively related to oscillations of the binding energy or the spectral weight. 
Experimentally, FDARPES was recently discussed in literature \cite{Hein2020,Suzuki2021,Lee2023,Ren2023} but a systematic, quantitative analysis has been lacking so far. 

Here, we present an in-depth study using frequency domain representations.
We set out by gaining intuition from a toy model before characterizing the experimentally observed coherent oscillations in WTe$_2$.
We evaluate two variations of frequency analysis by applying it to oscillations of (i) the photoemission intensity (PI)  and \tred{(ii) the first moment (FM) of the EDCs}. 
We extend earlier PI analysis\cite{Giovannini2020,DeGiovannini2022,Hein2020,Suzuki2021,Lee2023} with a systematic survey of applications and interpretations of experimental results. 
On the other hand, the FM analysis is novel and might provide a more intuitive interpretation. 
We begin in Sec.~\ref{sec:toymodel} by presenting both analysis methods for a simple, simulated dataset. 
In Sec.~\ref{ssec:COM} and \ref{ssec:Int}, we analyze binding energy oscillations by applying both methods to suitably complex trARPES data of the transition metal dichalcogenide WTe$_2$. 
Spectral weight oscillations in WTe$_2$ are examined in Sec.~\ref{ssec:SWosc}. 
In Sec.~\ref{ssec:Chirp}, we clarify how a linearly chirped probe pulse can lead to extrinsic artifacts in the frequency domain that are distinct from the intrinsic coherent phonon signal. 
We conclude in Sec.~\ref{sec:discussion} with a discussion of the advantages and limitations of both methods. 

%%%%%%%%%%%%%%%%%%%%%%%%%%%%%%%%%%%%%%%%%%%%%%%%%%%%%%%%
%%%%%%%%%%%%%%%%%%%%%%%%%%%%%%%%%%%%%%%%%%%%%%%%%%%%%%%%
\section{Introduction to frequency domain analysis}
\label{sec:toymodel}

trARPES measures the energy $E$ and momentum $k$ resolved photoemission intensity $I(k,E,t)$ as a function of pump-probe delay $t$. 
\tred{In the following, we assume equidistant delay steps of width $\delta_t$.}
To demonstrate the power of frequency domain analysis, we discuss the simplest model of a single Lorentzian peak in a single EDC $I(E,t)$. 
Insights of the exercise are easily generalized to more peaks and higher data dimensions.  
We will use these results to guide our analysis of experimental data in Sec.~\ref{sec:expresults}. 
The Lorentzian peak is centered at an energy $E=0$ with a full width at half maximum (FWHM) of 100~meV that oscillates with a frequency $f=2.4$~THz. 
We consider three scenarios for the origin of the oscillations and illustrate their transient EDCs in Fig.~\ref{fig:toymodel}a,f,k:
\begin{enumerate}
    \item oscillation of the binding energy $\Delta E=2$~meV,
    \item oscillation of the spectral weight $\Delta SW=2\%$,
    \item oscillation of the linewidth $\Delta W = 2$~meV.
\end{enumerate}
Binding energy oscillations originate from changing wave function overlap and electronic screening as ions move. 
Spectral weight oscillations are due to a change of orbital character and associated photoemission matrix elements. 
Linewidth oscillations can occur when the phase space for electron scattering process is modulated due to altered bands and their orbital character.

Our approach to frequency domain analysis is based on evaluating the Fourier component ${\cal F}_Q(k,E,f)$ of a quantity $Q(k,E,t) $ at a given frequency $f$,
\begin{equation}
    {\cal F}_Q^R(k,E,f)=\frac{2\delta_t}{t_N-t_1} \sum^{N}_{n=1} Q(k,E,t_n) \cos(2\pi f t_n),
\end{equation}
\begin{equation}
    {\cal F}_Q^I(k,E,f)=\frac{2\delta_t}{t_N-t_1} \sum^{N}_{n=1} Q(k,E,t_n) \sin(2\pi f t_n),
\end{equation}
where the indices $R$ and $I$ indicate the real and imaginary parts, respectively, and $\delta_t$ is the time bin size. 
For the sake of clarity, our model assumes pure cosine oscillations $\cos(2\pi f t)$ without a phase offset and therefore the imaginary Fourier component is zero.
We show below how each of the three scenarios produces a qualitatively different Fourier response using PI as well as FM analysis.

%%%%%%%%%%%%%%%%%%%%%%%%%%%%%%%%%%%%%%%%%%%%%%%%%%%%%%%%
%%%%%%%%%%%%%%%%%%%%%%%%%%%%%%%%%%%%%%%%%%%%%%%%%%%%%%%%
\subsection{Fourier analysis of photoemission intensity}
\label{sec:inttoymodel}

\begin{figure*}
\centering
\includegraphics[width=\textwidth]{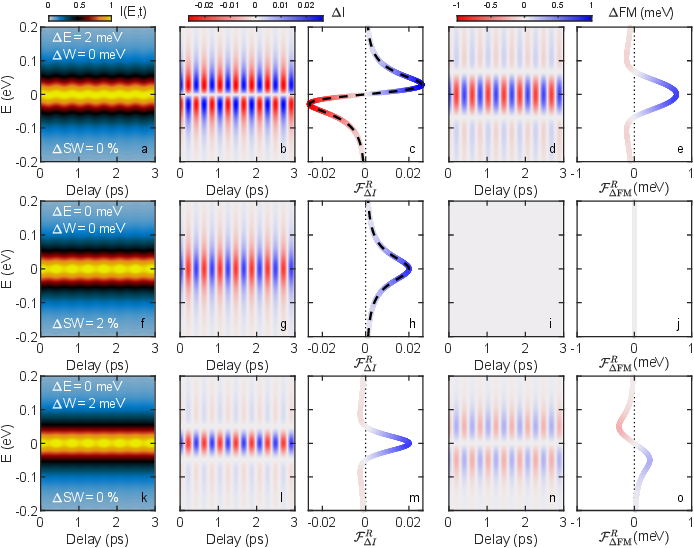}
\caption{PI and FM analysis of simulated 2.4~THz oscillation scenarios. 
(a)~Simulated PI $I(k,E,t)$ as a function of delay for a cosine oscillation of the binding energy with an amplitude $\Delta E=2$~meV. 
(b)~PI oscillations $\Delta I(E,t)$ after subtracting an energy-dependent, incoherent background. 
(c)~Real part of the 2.4~THz Fourier component of (b).
The black dashed line is given by Eq.~\ref{eq:BEoscillation}.
(d-e)~show the corresponding FM analysis. 
(d)~FM oscillations $\Delta \text{FM}(E,t)$ after subtracting an incoherent background.
A sliding integration window of 0.1~eV is taken at each delay to evaluate the FM at each energy.
(e)~Real part of the 2.4~THz Fourier component. 
(f-j)~PI and FM analysis for oscillations of the spectral weight $\Delta SW=2\%$. 
The black dashed line in panel (h) is given by Eq.~\ref{eq:SWoscillation}. 
(k-o)~PI and FM analysis for oscillations of the linewidth $\Delta W = 2$~meV.}
\label{fig:toymodel}
\end{figure*}

The first approach analyses PI oscillations $\Delta I(k,E,t)$ by subtracting a smooth, non-oscillatory (incoherent) $t$-dependent background at each $(k,E)$ pixel. 
In our model, this background is $t$-independent as we disregard relaxation processes. 
\tred{This allows unambiguous projection onto purely real Fourier components and highlights the characteristic structures of PI and FM analysis in Fig.~\ref{fig:toymodel}.}
$\Delta I(E,t)$ and its Fourier components ${\cal F}_{\Delta I}^R(E)$ at frequency $f=2.4$~THz are presented in Fig.~\ref{fig:toymodel}b,g,l and c,h,m, respectively.

\textbf{1. Binding energy oscillations:}
The largest signal occurs slightly above and below the peak maximum, without oscillations at the maximum (Fig.~\ref{fig:toymodel}b). 
The Fourier component in Fig.~\ref{fig:toymodel}c emphasizes a $\pi$ phase shift as the oscillation changes sign across the peak maximum. 
Previous theory work showed that the Fourier component is proportional to the energy derivative of the intensity\cite{Giovannini2020}:
\begin{equation}
    {\cal F}_{\Delta I}^R(E)=-\Delta E \times \frac{dI_\text{Ref}(E)}{dE},
    \label{eq:BEoscillation}
\end{equation} where the reference spectrum $I_\text{Ref}$ is taken before photo-excitation ($t<0$). 
The black dashed line in Fig.~\ref{fig:toymodel}c is the result of this relation for our model.
Eq.~\ref{eq:BEoscillation} is valid in the limit of an oscillation amplitude much smaller than the peak width, a condition that is often satisfied experimentally. 
The value of a binding energy oscillation $\Delta E$ can therefore directly be determined from an intensity analysis independent of specific line shapes. 

\tred{However, if bands overlap but have different oscillation amplitudes at the same frequency, the relation between the derivative and the Fourier component is lost as a linear superstition of responses according to Eq.~(\ref{eq:BEoscillation}) cannot be disentangled unambiguously.}
Furthermore, optical excitation creates non-oscillatory time dependencies of the spectral weight, binding energy and linewidth. 
These incoherent effects often dominate experimental data. 
Although the qualitative line shape of the Fourier component typically remains the same in theses cases, Eq.~\ref{eq:BEoscillation} is not exact since an appropriate reference spectrum does not exist. 
A suitable reference spectrum may be approximated by averaging the spectrum over the delay range used in the Fourier analysis. 
This is the strategy we pursue in the evaluation of our experimental data in Section~\ref{sec:expresults}. 

\textbf{2. Spectral weight oscillations:}
The largest signal occurs at the peak maximum (Fig.~\ref{fig:toymodel}g,h). 
As discussed in Ref.~\cite{Giovannini2020}, the Fourier component is proportional to the spectral function. 
\begin{equation}
  {\cal F}_{\Delta I}^R(E)=\Delta SW\times I_\text{Ref}(E).
      \label{eq:SWoscillation}
\end{equation}
This relation is shown by the black dashed line in Fig.~\ref{fig:toymodel}h.
As for the binding energy oscillations, exact equivalency is lost when non-oscillatory time dependencies are present.

\textbf{3. Width oscillations:}
The largest signal occur at the peak maximum (Fig.~\ref{fig:toymodel}l,m). 
Sign changes occur away from the peak maximum. 
A simple relation between the Fourier component and the width oscillations does not exist.

%%%%%%%%%%%%%%%%%%%%%%%%%%%%%%%%%%%%%%%%%%%%%%%%%%%%%%%%
%%%%%%%%%%%%%%%%%%%%%%%%%%%%%%%%%%%%%%%%%%%%%%%%%%%%%%%%
\subsection{Fourier analysis of \tred{the first moment of the EDCs}}
\label{sec:Comtoymodel}

The second approach is an analysis of the FM of the photoemission intensity. 
We define the FM as
\begin{equation}
    \text{FM}(k,E,t)=\frac{\int_{E-\delta/2}^{E+\delta/2} I(E',k,t) E' dE'}{\int_{E-\delta/2}^{E+\delta/2} I(E',k,t) dE'},
\end{equation}
where $\delta$ is the energy integration window size.
We discuss its influence in more detail below. 
$(k,E)$ pixel-by-pixel analysis is implemented with a sliding energy window of width $\delta$.
As in the PI analysis, the oscillating part of the FM is obtained by subtracting an incoherent, $t$-dependent background at each pixel.
The resulting oscillatory signals $\Delta \text{FM} (E,t)$ and its Fourier components ${\cal F}_{\Delta FM}^R (E)$ are presented in Fig.~\ref{fig:toymodel}d,i,n and Fig.~\ref{fig:toymodel}e,j,o, respectively. 
The integration window $\delta$ is 0.1~eV in these examples.

\begin{figure}[t]
\centering
\includegraphics[width=\columnwidth]{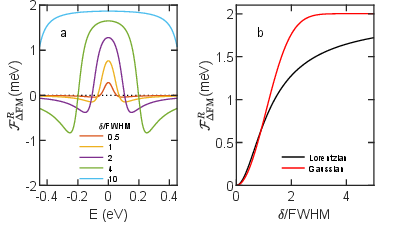}
\caption{(a) Fourier component of the FM oscillations as function of integration window width $\delta$ for identical oscillation parameters as in Fig.~\ref{fig:toymodel}e. 
(b) Fourier component at $E=0$~eV as function of $\delta$ for Lorentzian or Gaussian line shape.}
\label{fig:IntegrationWindow}
\end{figure}

\textbf{1. Binding energy oscillations:}
FM oscillations are strongest around the peak maximum (Fig.~\ref{fig:toymodel}d,e). 
The Fourier amplitude ${\cal F}_{\Delta FM}^R$ at the peak maximum is directly related to the binding energy oscillations $\Delta E$. 
For the parameters here, the maximum of ${\cal F}_{\Delta FM}^R$ is slightly less than 1~meV while $\Delta E= 2$~meV. 
This discrepancy originates from the finite size of the energy integration window as illustrated in Fig.~\ref{fig:IntegrationWindow}a. 
Increasing the integration window leads to a larger Fourier component at the peak position.
We trace this dependency for Lorentzian and Gaussian line shapes in Fig.~\ref{fig:IntegrationWindow}b.
For a Gaussian peak, the integration window must be at least two times larger than the Gaussian FWHM before ${\cal F}_{\Delta FM}^R=\Delta E$. 
For a Lorentzian peak, ${\cal F}_{\Delta FM}^R$ approaches $\Delta E$ more slowly. 
While this suggests to maximize the integration window,  real datasets often require $\delta$ to be small enough to avoid averaging multiple bands. 
Practically, results from ${\cal F}_{\Delta FM}^R$ will almost always underestimate $\Delta E$. 
Furthermore, it is straightforward to show that a constant background also reduces FM oscillations.

\textbf{2. Spectral weight oscillations:}
The FM is not affected by spectral weight oscillations and the corresponding Fourier component is exactly zero everywhere (Fig.~\ref{fig:toymodel}i,j).

\textbf{3. Width oscillations:}
The largest FM oscillations occur slightly above and below the peak maximum, without oscillations at the peak maximum (Fig.~\ref{fig:toymodel}n). The Fourier component in Fig.~\ref{fig:toymodel}o emphasizes a change of sign above and below the maximum. 

%%%%%%%%%%%%%%%%%%%%%%%%%%%%%%%%%%%%%%%%%%%%%%%%%%%%%%%%
%%%%%%%%%%%%%%%%%%%%%%%%%%%%%%%%%%%%%%%%%%%%%%%%%%%%%%%%
\subsection{Comparison of PI and FM analysis}

Both PI and FM analysis can distinguish characteristic signatures of different types of oscillations. 
PI analysis is sensitive to binding energy, intensity and width oscillations.
It can quantitatively relate the Fourier component to the amplitude of binding energy oscillations. 
A physically meaningful interpretation requires careful consideration of this relationship since the size of the Fourier components on its own does not signal large energy oscillation amplitudes. 
For example, bands with large spectral weight or narrow linewidth have large Fourier components even with weak binding energy oscillations. 
Indeed, the sharp line of the surface state in Gadolinium enabled the first observation of a coherent phonon in a metal with trARPES \cite{Loukakos2007}.
On the other hand, the FM analysis is constructed to emphasize oscillations of the binding energy at the band position and hence is often easier to interpret. 
The size of the FM Fourier component is directly related to the amplitude of the binding energy oscillations but systematically underestimated.
The FM provides a lower-bound estimate of binding energy oscillations as long as the energy integration window is comparable or larger than the linewidth.
Both approaches have the advantage of being independent of the line shape.
However, PI analysis cannot correctly quantify binding energy oscillations in overlapping bands and FM analysis likewise fails to isolate independent oscillations of overlapping bands.
\tred{The combination of both methods can distinguish width from intensity oscillations as the latter are absent in FM analysis.
This selectivity may be useful when investigating non-linear electron-phonon coupling \cite{Li2013,DeGiovannini2022} or expected coherent responses of quantum condensates \cite{Kemper2015,Nosarzewski2017,Rustagi2019}.}

%%%%%%%%%%%%%%%%%%%%%%%%%%%%%%%%%%%%%%%%%%%%%%%%%%%%%%%%
%%%%%%%%%%%%%%%%%%%%%%%%%%%%%%%%%%%%%%%%%%%%%%%%%%%%%%%%
\section{Experimental details}
\label{sec:expdetails}
Our experimental trARPES setup uses a Ti:Sapphire regenerative amplifier which operates at a repetition rate of 312~kHz and emits 1.5~eV photons to generate pump and probe pulses. 
6~eV probe pulses are obtained through two stages of second harmonic generation in $\beta$-BaB$_2$O$_4$ non-linear crystals\cite{Gauthier2020a}. 
We use two different pulse width settings for the 1.5~eV pump pulse: 
The fast setting uses the 35~fs pump pulses as generated by the amplifier. 
In the slow setting, 5~cm of SF11 glass are added to the pump path to stretch the pulse. 
The overall time resolution is estimated 130~fs and 600~fs in the fast and slow settings, respectively. 
The slow setting strongly suppresses higher frequency coherent phonons and aids studies of lower frequency modes. 
\tred{}
Data can be recorded over a larger delay range with a lower sampling rate without artifacts due to beating of low and high frequency modes.
The beam profiles were $91 \times 65$~$\mu$m$^2$ for the pump and $34 \times 33$~$\mu$m$^2$ for the probe.
The incident pump excitation density was 0.28~mJ/cm$^2$ in the fast setting and 0.70~mJ/cm$^2$ in the slow setting. 
\tred{Pump pulses were $p$-polarized while probe pulse were either $p$ or $s$ polarized.}
Photoelectrons were detected with a hemispherical analyzer operating in \tred{dither mode \cite{Sobota2021,Liu2022}} at a pass energy of 10~eV. 
The sample was biased at -25~V to access a wider momentum range \cite{Gauthier2021a}.

Commercial WTe$_2$ samples were obtained from HQ Graphene. 
The sample was cleaved in situ at a base pressure of $<1\times 10^{-10}$~Torr and kept at a temperature of 80~K. 
All data was obtained from a single cleave and are representative of typical results obtained on this material \cite{Hein2020}. 
The absence of inversion symmetry in WTe$_2$ allows for two distinct cleaved surfaces.
Measurements were performed along the $\overline{\Gamma}-\overline{X}$ direction on a type A surface \cite{Bruno2016} with a work function of 5.05~eV. 
The long acquisition times required for high quality data can lead to systematic drifts in energy, momentum, and time, which we corrected \cite{Sakamoto2022}.

%%%%%%%%%%%%%%%%%%%%%%%%%%%%%%%%%%%%%%%%%%%%%%%%%%%%%%%%
%%%%%%%%%%%%%%%%%%%%%%%%%%%%%%%%%%%%%%%%%%%%%%%%%%%%%%%%
\section{Experimental results}
\label{sec:expresults}

\begin{figure}
\centering
\includegraphics[width=\columnwidth]{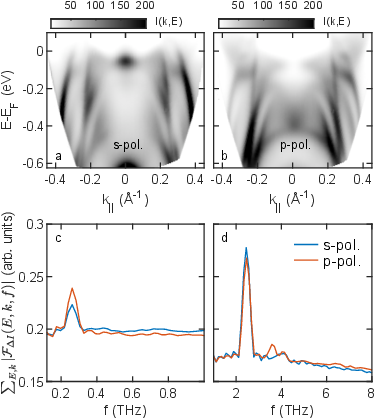}
\caption{(a-b) Time-averaged intensity $I_\text{Ref}(k,E)$ for $s$ and $p$ polarized probe.
Fourier transform of PI oscillations integrated over $(k,E)$ for $s$ and $p$ probe polarization and for (c) the slow and (d) the fast setting.}
\label{fig:determineFreq}
\end{figure}

We begin by presenting an overview of the time-averaged trARPES data of WTe$_2$ in Fig.~\ref{fig:determineFreq}a-b for $s$ and $p$ probe polarization taken in the fast setting.
We analyze oscillations in a delay range of  0.18-3.1~ps for the fast setting and 5-20~ps for the slow setting. 
The intensity $I_\text{Ref}(k,E)$ is averaged over the same delay ranges.
$I_\text{Ref}(k,E)$ approximates a reference spectrum for further analysis and minimizes discrepancies from photoinduced binding energy shifts.
\tred{In the following figures, we suppress the index and designate the reference spectrum as $I(k,E)$.}
We observe a multitude of bands and their distinct orbital characters are highlighted by the polarization dependence.
In the following, we will show that 4 optical phonons couple to all bands with varying degree and attempt to quantify binding energy oscillations of the two strongest modes.

First, we identify relevant frequencies in Fig.~\ref{fig:determineFreq}c-d.
We apply the PI analysis to each pixel and integrate its absolute value over the complete spectrum. 
This approach requires a low noise level to avoid accumulating a large frequency-independent background.
In the slow setting in Fig.~\ref{fig:determineFreq}c, we observe one mode at 0.26~THz for both $s$ and $p$ probe polarizations.
In the fast setting in Fig.~\ref{fig:determineFreq}d, we observe three modes 2.45, 3.63 and 4.03~THz. 
\tred{In the slow setting, the pump pulse only excites the 0.26~THz mode coherently as no wave packet can form for the faster modes. 
Conversely, in the fast setting, we sample less than a full period of the 0.26~THz mode and any component at this frequency becomes subject to background subtraction.}

The precise frequency values were determined by the fits presented in Appendix~\ref{sec:appTimeZero}, Fig.~\ref{fig:determinet0}c,d.
The 2.45~THz mode dominates the Fourier spectrum and the 3.63~THz mode is only present for a $p$-polarized probe. 
These frequencies agree well with $A_1$ phonon modes in WTe$_2$ as reported by Raman scattering~\cite{Ma2016} and previous pump-probe experiments~\cite{He2016,Sie2019, Soranzio2019,Hein2020,Drueke2021,Ji2021a,Soranzio2022}.
For clarity, we focus on the two strongest modes at 0.26~THz and 2.45~THz for $p$-polarized spectra.
A complete set of PI and FM analysis of all four modes from $s$- and $p$-polarized spectra is given in Appendix~\ref{sec:appFDARPES}.

In the spirit of our toy model in Section~\ref{sec:toymodel}, we attempt to choose an effective $t_0$ such that the imaginary (real) part of the Fourier component is zero and all information is contained in the real (imaginary) part.
We are indeed able to chose a global, effective $t_0$ to achieve this decomposition. 
The details of this $t_0$ determination are given in Appendix~\ref{sec:appTimeZero}.
The decomposition into purely real or imaginary parts does not always succeed for our experimental data. 
We discuss the possible implications of this observation in Section~\ref{sec:discussion}.
Our phenomenologically defined, effective $t_0$ includes the true time-zero and mode-dependent phase shifts and allows discussing the phase difference of modes but it does not reveal their true phase.
This data treatment thus does not provide reliable information of excitation mechanisms \cite{Merlin1997} or electronic driving forces~\cite{Li2013}.
In the following, we focus on isolating and quantifying electronic oscillation amplitudes in a complex band structure.
To this end, we develop composite FDARPES maps for a comprehensive yet intuitive visual representation of the absolute energy oscillation amplitudes. 

%%%%%%%%%%%%%%%%%%%%%%%%%%%%%%%%%%%%%%%%%%%%%%%%%%%%%%%%
%%%%%%%%%%%%%%%%%%%%%%%%%%%%%%%%%%%%%%%%%%%%%%%%%%%%%%%%
\subsection{Binding energy oscillations from FM analysis}
\label{ssec:COM}

\begin{figure}
\centering
\includegraphics[width=\columnwidth]{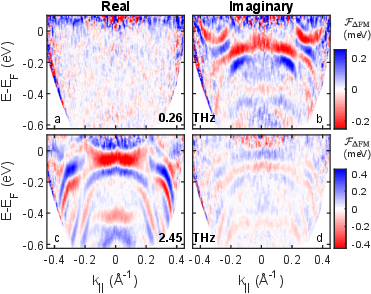}
\caption{FM analysis of binding energy oscillations for $p$-polarized spectra of the 0.26~THz (a-b) and 2.45~THz modes (c-d).
The real (imaginary) Fourier components are presented in the left (right).}
\label{fig:FmapCOMppol}
\end{figure}

We show the result of the FM analysis in Fig.~\ref{fig:FmapCOMppol} where we used an energy integration window of $\delta=0.1$~eV.
\tred{The integration window is larger than the observed peak widths to ensure that the characteristic oscillatory structures are robustly captured.}
We observe a rich coherent response in a multitude of bands.
The 0.26~THz mode projects entirely on the imaginary axis while the 2.45~THz mode is largely real.
Both modes have a $\pi/2$ phase difference but we cannot tell their true phase due to our quasi-free choice of $t_0$.
All essential information is either contained in the real or imaginary Fourier component while the complement is largely blank. 
However, weak imaginary signatures are present for the 2.45~THz mode in Fig.~\ref{fig:FmapCOMppol}d, see also Appendix~\ref{sec:appFDARPES}.

\begin{figure}
\centering
\includegraphics[width=\columnwidth]{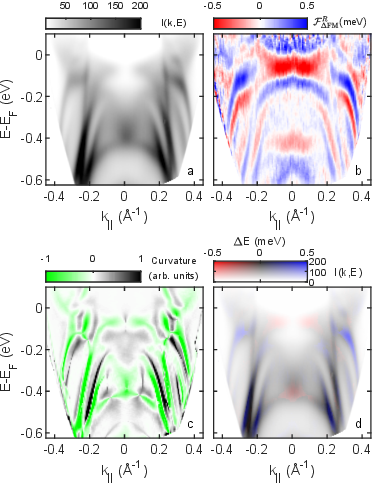}
\caption{(a)~Time-averaged intensity $I_\text{Ref}$ as in Fig.~\ref{fig:determineFreq}a. 
(b)~2.45~THz FM oscillations as in Fig.~\ref{fig:FmapCOMppol}d. 
(c)~Negative (green) values of the two-dimensional curvature highlight bands in (a).  
(d)~Composite FDARPES spectrum of the band oscillations resulting from the 2.45~THz phonon mode. 
A two-dimensional colormap projects the band oscillations ${\cal F}_{\Delta \text{FM}}^R$ on the spectrum $I_\text{Ref}$. 
The curvature is used as a mask, see text.}
\label{fig:ExampleCOMppol}
\end{figure}

The frequency domain representations of the FM oscillations in Fig.~\ref{fig:FmapCOMppol} and Appendix~\ref{sec:appFDARPES} are rich in information but not straightforward to interpret. 
Much of the band structure oscillates but a direct relation to the band intensities (density of states) is not obvious.
In Sec.~\ref{sec:Comtoymodel}, we realized that the FM Fourier component at the band position directly corresponds to the amplitude of binding energy oscillations $\Delta E$.
Here, we introduce a method to construct more intuitive composite FDARPES spectra to visualize $\Delta E$ directly on the bands by combining the Fourier component with the photoemission intensity.
Fig.~\ref{fig:ExampleCOMppol}d shows the composite FDARPES spectrum of the 2.45~THz mode.

First, we use the time-averaged intensity $I_\text{Ref}(k,E)$ in Fig.~\ref{fig:ExampleCOMppol}a to define a mask that selects Fourier components (Fig.~\ref{fig:ExampleCOMppol}b) which are located close to the band positions. 
The two-dimensional curvature of the photoemission intensity \cite{Zhang2011} in Fig.~\ref{fig:ExampleCOMppol}c highlights the band dispersions in its negative (green) parts which serve as a mask.
As we showed in Sec.~\ref{sec:Comtoymodel}, Fourier components far from the band positions are strongly dependent on the width of the energy integration window and thus less physically relevant.

Second, we combine the masked Fourier spectrum with the photoemission intensity in a two-dimensional colormap (Fig.~\ref{fig:ExampleCOMppol}d), a visualization commonly used in the presentation of multi-dimensional spin-resolved ARPES data \cite{Tusche2015,Jozwiak2016,Lee2023}. 
One color axis describes the photoemission intensity while the second color axis describes the oscillation amplitudes $\Delta E$. 
The composite FDARPES spectra offer an intuitive overview of the orbital-dependent electronic oscillation amplitudes of a specific phonon mode. 
For the example of a $p$-polarized probe in Fig.~\ref{fig:ExampleCOMppol}d, oscillations of the 2.45~THz mode are localized in 4 regions of the band structure and amplitudes are smaller than 0.5~meV.
We show composite FDARPES spectra for all modes and probe polarizations in Appendix~\ref{sec:appFDARPES}.

\begin{figure}
\centering
\includegraphics[width=\columnwidth]{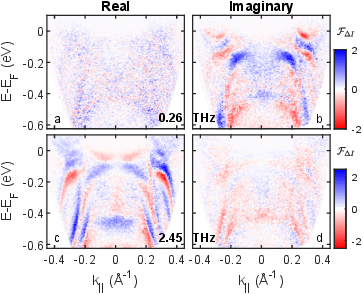}
\caption{PI analysis of binding energy oscillations for $p$-polarized spectra of the 0.26~THz (a-b) and 2.45~THz modes (c-d).
The real (imaginary) Fourier components are presented in the left (right).}
\label{fig:FmapIntppol}
\end{figure}

%%%%%%%%%%%%%%%%%%%%%%%%%%%%%%%%%%%%%%%%%%%%%%%%%%%%%%%%
%%%%%%%%%%%%%%%%%%%%%%%%%%%%%%%%%%%%%%%%%%%%%%%%%%%%%%%%
\subsection{Binding energy oscillations from PI analysis}
\label{ssec:Int}

\begin{figure}
\centering
\includegraphics[width=0.9\columnwidth]{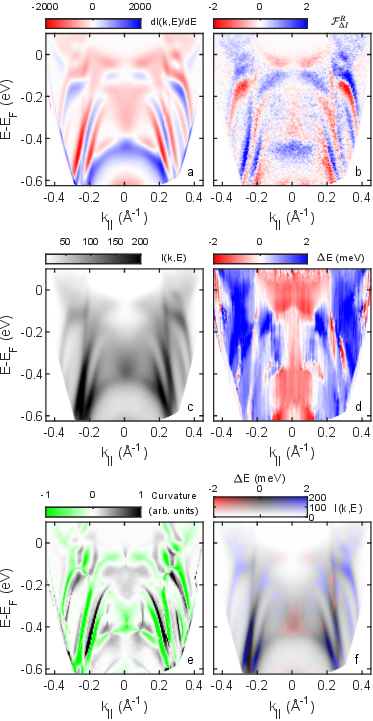}
\caption{(a)~Energy derivative $dI_\text{Ref}/dE$ of (c). 
(b)~PI oscillations at 2.45~THz as in Fig.\ref{fig:FmapIntppol}c. 
(c)~Time-averaged intensity $I_\text{Ref}$ as in Fig.~\ref{fig:determineFreq}a. 
\tred{(d)~Energy amplitude $\Delta E(f=2.45\text{~THz})$ by fitting the prefactor in eq.~(\ref{eq:BEoscillation}) to the energy derivative in (a) and the frequency domain representation in (b), see text.} 
(e)~Negative (green) values of the two-dimensional curvature highlight bands in (c). 
(f)~Composite FDARPES spectrum of the band oscillations resulting from the 2.45~THz phonon mode. 
A two-dimensional colormap projects the band oscillations $\Delta E(f=2.45\text{~THz})$ on the spectrum $I_\text{Ref}$. 
The curvature is used as a mask, see text.}
\label{fig:exampleInt}
\end{figure}

We show the result of the PI analysis in Fig.~\ref{fig:FmapIntppol}.
As for the FM analysis, decomposition of real and imaginary Fourier components is well defined, see also Appendix~\ref{sec:appFDARPES}.
Sec.~\ref{sec:inttoymodel} illustrated how binding energy oscillations generate Fourier components that are proportional to the energy derivative of the spectrum. 
The proportionality factor is the oscillation amplitude $\Delta E$ (Eq.~\ref{eq:BEoscillation}). 
We therefore compare ${\cal F}_{\Delta I}(k,E,f)$ and ${dI_\text{Ref}(k,E)}/{dE}$ side by side in Fig.~\ref{fig:exampleInt}a and b. 
Both quantities share many similarities, which indicates the dominance of binding energy oscillations.
We find a similar correspondence for all polarizations and frequencies, except for the 0.26~THz mode, which we discuss in detail in Sec.~\ref{ssec:SWosc}. 

Motivated by these similarities, we construct composite FDARPES spectra in Fig.~\ref{fig:exampleInt}, similar to the FM analysis.
First, assuming that ${\cal F}_{\Delta I}(k,E,f)$ originates solely from binding energy oscillations, we use a sliding window of 0.1~eV to fit the proportionality factor at every pixel in (Fig.~\ref{fig:exampleInt}d). 
This quantity corresponds to $\Delta E$ near the bands; other regions are not physically relevant. 
As before, we define a mask using the two-dimensional curvature (Fig.~\ref{fig:exampleInt}e) to select the relevant regions. 
Finally, we combine the masked $\Delta E$ with the photoemission intensity into a composite FDARPES spectrum in Fig.~\ref{fig:exampleInt}f. 
Composite FDARPES spectra for all modes and probe polarizations are shown in Appendix~\ref{sec:appFDARPES}.
The result of PI analysis is qualitatively similar to the FM analysis in Figs.~\ref{fig:ExampleCOMppol}d.
Absolute amplitude values range up to 2~meV.
This is $\sim$~4 times larger than in FM analysis and wider sections of the band structure exhibit considerable oscillation amplitudes.
We compare both methods in more detail in Sec.~\ref{sec:discussion}.

%%%%%%%%%%%%%%%%%%%%%%%%%%%%%%%%%%%%%%%%%%%%%%%%%%%%%%%%
%%%%%%%%%%%%%%%%%%%%%%%%%%%%%%%%%%%%%%%%%%%%%%%%%%%%%%%%
\subsection{Spectral weight oscillations from PI analysis}
\label{ssec:SWosc}

\begin{figure}
\centering
\includegraphics[width=\columnwidth]{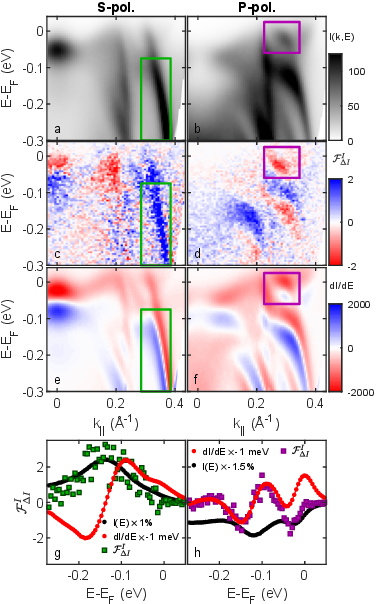}
\caption{(a-b)~Time-averaged intensity $I_\text{Ref}$ for $s$- and $p$-polarization.
(c-d)~PI analysis of the 0.26~THz mode. 
(e-f)~Energy derivative $dI_\text{Ref}/dE$. 
Colored boxes in (a-f) indicate regions of interest discussed in the text. 
(g)~Comparison of EDCs of ${\cal F}_{\Delta I}^{I}(f=0.26\text{~THz})$, $I_\text{Ref}$ and $dI_\text{Ref}/dE$ with $s$ polarization for $k_{||}=0.335$~\AA$^{-1}$ (central value of the green box). 
(h)~Same as (g) with $p$ polarization for $k_{||}=0.285$~\AA$^{-1}$ (central value of the purple box).}
\label{fig:13BeyondBEosci}
\end{figure}

Binding energy oscillations due to coherent phonons are readily understood by a periodic change of overlap integrals.
Accordingly, frozen-phonon density functional theory reproduces experimental findings well in semiconductors and semimetals \cite{Papalazarou2012,Faure2013,Sobota2014,Golias2016,Sakamoto2022}.
Most oscillations we observe in WTe$_2$ can indeed be attributed to binding energy oscillations. 
In contrast, oscillations of the spectral weight and linewidth are scarcely addressed in literature. 
In this section, we use the PI analysis to reveal spectral weight oscillations of the 0.26~THz mode.
\tred{We speculate that spectral weight oscillations could be due to a change of orbital
character and associated photoemission matrix elements.}

Our model in Sec.~\ref{sec:inttoymodel} informed us that spectral weight oscillations result in Fourier components that are proportional to the intensity $I_\text{Ref}(k,E)$ (eq.~\ref{eq:SWoscillation}) rather than its energy derivative.
Accordingly, in Fig.~\ref{fig:13BeyondBEosci}, we compare $I_\text{Ref}(k,E)$(a-b) to the PI analysis (c-d) and energy derivative (e-f).
Two regions show spectral weight oscillations most clearly as they satisfy eq.~\ref{eq:SWoscillation} and are highlighted with boxes. 
To quantify this effect, we compare EDCs of the PI analysis, the spectrum and its energy derivative in Fig.~\ref{fig:13BeyondBEosci}g,h. 
${\cal F}_{\Delta I}(k,E,f)$ inside the green box is proportional to $I_\text{Ref}(k,E)$ rather than ${dI_\text{Ref}}/{dE}(k,E)$ (Fig.~\ref{fig:13BeyondBEosci}g). 
The spectral weight oscillation has an amplitude of 1\%. 
Equivalently, we find a spectral weight oscillation with an amplitude of 1.5\% close to the Fermi level inside the purple box  (Fig.~\ref{fig:13BeyondBEosci}h). 
\tred{At binding energies larger than 0.05~eV, the oscillation type changes and oscillations are well described by binding energy oscillations with an amplitude of 1 meV.}

%%%%%%%%%%%%%%%%%%%%%%%%%%%%%%%%%%%%%%%%%%%%%%%%%%%%%%%%
%%%%%%%%%%%%%%%%%%%%%%%%%%%%%%%%%%%%%%%%%%%%%%%%%%%%%%%%
\subsection{Effects of a linearly chirped probe pulse}
\label{ssec:Chirp}

In Sec. \ref{sec:expresults} and Appendix~\ref{sec:appTimeZero}, we chose an effective $t_0$ such that all oscillations are described by the real (or imaginary) Fourier component. 
However, in some cases, weak features remained in the complimentary component which indicates the presence of oscillations with a phase shift of $\pm\pi/2$ relative to all the other oscillations. 
We will show in the following that those $\pm\pi/2$ out-of-phase oscillations are artefacts caused by the convolution of a linearly chirped probe pulse with the time-dependent spectral intensity. 

To illustrate this, we performed a numerical two-dimensional convolution of a linearly chirped probe pulse with a simulated spectral intensity $S(E,t)$. 
The simulated spectral intensity represents a Gaussian peak with FWHM of $\Gamma$ and binding energy oscillations $\Delta E= 0.1 \Gamma$ at frequency $\Omega$, as shown in Fig.~\ref{fig:14Chirp}c. 
The Wigner distribution of a linearly chirped pulse (see Appendix \ref{sec:appMathChirp}, Eq.~\ref{eq:Wigner}) is illustrated in Fig.~\ref{fig:14Chirp}b, in units of $\Gamma$ and $T=2\pi/\Omega$. 

\begin{figure}
\centering
\includegraphics[width=\columnwidth]{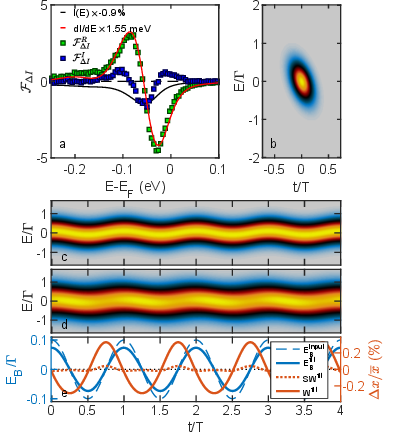}
\caption{(a)~EDCs at $k_{||}=0$ of ${\cal F}_{\Delta I}^{R/I}(f=2.45\text{~THz})$, $I_\text{Ref}$ and $dI_\text{Ref}/dE$ in $s$ polarization. 
(b)~Wigner distribution of a linearly chirped pulse with $\eta=0.5$, in units of the band parameters in (c).
(c)~Simulated spectral intensity of Gaussian peak with FWHM $\Gamma$.
The band energy oscillates with an amplitude $\Delta E= 0.1 \Gamma$ at a frequency $1/T$.  
(d)~Convolution of the simulated spectral intensity in (c) with the Wigner distribution of the linearly chirped pulse in (b). 
(e)~Fitted binding energy, linewidth and spectral weight of the convoluted spectral intensity. 
The linewidth oscillates $-\pi/2$ out of phase relative to the binding energy.}
\label{fig:14Chirp}
\end{figure}

The resulting convolution in Fig.~\ref{fig:14Chirp}d shows the intensity oscillating $\pm\pi/2$ out of phase of the binding energy oscillations. 
We fit the simulated data set with a Gaussian function and obtain the  parameters in Fig.~\ref{fig:14Chirp}e. 
The band binding energy $E_B^\text{fit}$ retains its original cosine behaviour, with a slightly reduced amplitude relative to the intrinsic value $E_B^\text{input}$, as is the case in absence of chirp. 
The total spectral weight $SW^\text{fit}$ has a negligible time-dependence and is not strongly affected by weak linear chirp. 
The linewidth $W^\text{fit}$ is most affected by linear chirp and exhibits sine-like oscillations. 

More rigorous mathematical arguments are presented in Appendix~\ref{sec:appMathChirp} and support the conclusion of our numerical model. 
In short, a linearly chirped pulse can be approximated by the sum of an even function (no chirp) and an odd function (linear chirp). 
The convolution with the time-dependent spectral intensity transmits   odd and even characters to the measured intensity. 
In consequence, to first order, linear chirp induces extrinsic oscillations that are $\pm\pi/2$ out-of-phase from the intrinsic oscillations. 
This out-of-phase component can lead to effective phase shifts in experiments and requires careful checking during data analysis. 
Fortunately, as the numerical simulation demonstrates, intrinsic and extrinsic oscillations are of different types and can be distinguished. 

This numerical simulation helps to understand our experimental results. 
We observe the strongest signal in the imaginary part of the Fourier component for the $2.45$~THz mode at $k_{||}=0$ near the Fermi level ($E_F$) for the $s$-polarized spectrum (Fig.~\ref{figApp:FmapIntspol}d in Appendix~\ref{sec:appFDARPES}). 
In Fig.~\ref{fig:14Chirp}a, we compare EDCs at $k_{||}=0$ of the Fourier components ${\cal F}_{\Delta I}^{R/I}$ to the intensity $I$ and its derivative $dI/dE$. 
The real part ${\cal F}_{\Delta I}^{R}$ is well described by $dI_\text{Ref}/dE$, indicating binding energy oscillations with an amplitude of 1.55~meV. 
In contrast, the line shape of the imaginary part ${\cal F}_{\Delta I}^{I}$ differs from both $I$ and $dI/dE$, indicating that it is not caused by oscillations of binding energy or spectral weight. 
The line shape is similar to the one expected for width oscillations (Fig.~\ref{fig:toymodel}m). 
We therefore conclude that the weak features in the complements of the Fourier components are linewidth oscillations with a phase shift of $\pm\pi/2$ relative to binding energy oscillations.
By comparison with the numerical simulation, we attribute their origin to the presence of a small but finite linear chirp of the probe pulse in our experiment. 
Based on the simulation and experimental results, we estimate that the probe pulse had a group delay dispersion of $\sim 3000$~fs$^2$.
While this value may seem large it is plausible as it corresponds to stretching a 100~fs pulse to 130~fs, our experimentally observed time resolution.

%%%%%%%%%%%%%%%%%%%%%%%%%%%%%%%%%%%%%%%%%%%%%%%%%%%%%%%%
%%%%%%%%%%%%%%%%%%%%%%%%%%%%%%%%%%%%%%%%%%%%%%%%%%%%%%%%
\section{Discussion and conclusions}
\label{sec:discussion}

\begin{figure}
\centering
\includegraphics[width=\columnwidth]{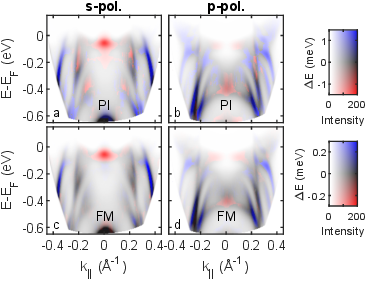}
\caption{(a-b)~Composite $s$ and $p$-polarized FDARPES spectra from PI analysis of the 2.45~THz mode. 
(c-d)~Corresponding composite FDARPES spectra using FM analysis.}
\label{fig:CompositeCOM}
\end{figure}

Fig.~\ref{fig:CompositeCOM} compares composite FDARPES spectra from FM and PI analysis for the strongest mode at 2.45~THz.
The orbital-dependent nature of the bands near the $\Gamma$-point is immediately evident by the pronounced polarization dependence.
Both numerical methods are well suited for the analysis of multiple coherent phonons in complex band structures as they avoid fitting analytical line shapes.
Both can quantify binding energy oscillations, the most common type of electronic coherent response in trARPES.
They give qualitatively similar results, but differ quantitatively as each has its advantages and limitations.

The FM Fourier component at a band position corresponds directly to the amplitude of its binding energy oscillations. 
The FM analysis is insensitive to spectral weight and linewidth oscillations at the band position (Fig.~\ref{fig:toymodel}) and therefore provides a well-defined quantity with straightforward interpretation. 
However, oscillation amplitudes are systematically underestimated due to the finite energy integration window (Fig.~\ref{fig:IntegrationWindow}) and electronic relaxation processes. 
As an example, the FM oscillation amplitude for the 2.45~THz mode at $k_{||}=0$ in $s$ polarization is $\sim0.7$~meV, while conventional peak fitting results in an oscillation amplitude of 1.58~meV. 
When also considering an exponential amplitude decay in the fits, the amplitude is 2.05~meV at $t_0$.
FM analysis cannot distinguish overlapping bands, yet it still provides a meaningful assessment of their shared behaviour. 

The strength of the PI analysis is its ability to distinguish oscillations in binding energy, line width and spectral weight by their characteristic signatures in the frequency domain (Fig.~\ref{fig:toymodel}).  
These signatures are directly related to the spectrum itself (Eqs.~\ref{eq:BEoscillation},\ref{eq:SWoscillation}), guaranteeing the applicability for any spectral line shapes. 
The completeness of the PI analysis also implies that its interpretation is more difficult as oscillations of mixed physical origin or of overlapping bands with different responses can lead to ambiguous features. 
In simple cases, the PI analysis estimates binding energy oscillations more accurately than the FM approach. 
As an example, the oscillation amplitude for the 2.45~THz mode at $k_{||}=0$ in $s$ polarization is 1.55~meV as compared to 1.58~meV from conventional fits without exponential decay. 
However, common incoherent relaxation processes can lead to underestimation of the amplitudes.
Results are best regarded as lower bound estimates, also when effects of finite time resolution are considered \cite{Gerber2017}.

Both approaches require that oscillations do not overlap in the frequency domain to avoid interference effects. 
Decaying oscillation amplitudes lead to a finite linewidth in the frequency domain and a frequency-dependent complex phase.
Consequently, rapidly decaying oscillations are more likely to overlap in the frequency domain and the imaginary part of the Fourier component could remain finite.
We consider the observed modes in WTe$_2$ well enough separated and did not account for such effects.

It is not always possible to decompose a signal into purely real or imaginary components, see for example Fig.~\ref{fig:FmapCOMppol}d.
\tred{Several explanations are possible: trivial sample physics includes phonon damping and interference between neighboring modes.
Trivial extrinsic effects may be due to imperfect background subtraction and a linear chirp of the probe pulse induces as we demonstrate here.}
In our experiment, a linear chirp of the probe pulse generated spurious Fourier components that were $\pm\pi/2$ out-of-phase from the intrinsic, physically meaningful oscillations.
This might be a common artifact due to the challenge to resolve the detailed time structure of the probe pulses in trARPES and mitigate such effects effectively.
Non-trivial explanations of the impossibility to define a unique phase which globally decomposes the Fourier components include non-adiabatic electron-phonon coupling \cite{DeGiovannini2022} and decay of the electronic driving forces \cite{Li2013}.
We exclude these explanations for our observations in WTe$_2$.
As our work shows, subtle, experimental effects influence the analysis of the phase and we encourage careful, methodical checks for possible artifacts.

In summary, frequency domain approaches are generally applicable to study coherent oscillations in any trARPES datasets. 
The number of free variables is small and easily adapted to different data sets:
only the delay range, relevant frequencies, effective $t_0$, the order of the polynomial background, and the width of the sliding window need to be adjusted.
Accordingly, the FDARPES analysis can be largely automatized to provide a complete overview of the coherent response, which can guide a more detailed quantitative analysis of selected features. 
FDARPES analysis is well suited to characterize the coherent response in increasingly large datasets inherent to trARPES measurements and guide discussion toward the most interesting physics.  

%%%%%%%%%%%%%%%%%%%%%%%%%%%%%%%%%%%%%%%%%%%%%%%%%%%%%%%%
%%%%%%%%%%%%%%%%%%%%%%%%%%%%%%%%%%%%%%%%%%%%%%%%%%%%%%%%
\begin{acknowledgments}
This work was supported by the U.S. Department of Energy, Office of Basic Energy Sciences, Materials Science and Engineering Division under Award DE-AC02-76SF00515. 
N.G. acknowledges support from the Swiss National Science Foundation (Fellowship No. P2EZP2 178542).
H.P. is supported by the U.S. Department of Energy, Office of Basic Energy Sciences, Materials Sciences and Engineering Division under Award DE-SC0024135.
\end{acknowledgments}
%%%%%%%%%%%%%%%%%%%%%%%%%%%%%%%%%%%%%%%%%%%%%%%%%%%%%%%%
%%%%%%%%%%%%%%%%%%%%%%%%%%%%%%%%%%%%%%%%%%%%%%%%%%%%%%%%

%%%%%%%%%%%%%%%%%%%%%%%%%%%%%%%%%%%%%%%%%%%%%%%%%%%%%%%%
%%%%%%%%%%%%%%%%%%%%%%%%%%%%%%%%%%%%%%%%%%%%%%%%%%%%%%%%
\section*{Data Availability Statement}
All 4 trARPES data sets on which this work is based (fast and slow setting in $s$ and $p$ probe polarization) are openly available in Stanford Digital Repository \url{https://purl.stanford.edu/vw292dt0072} at \url{https://doi.org/10.25740/vw292dt0072}.

%%%%%%%%%%%%%%%%%%%%%%%%%%%%%%%%%%%%%%%%%%%%%%%%%%%%%%%%
%%%%%%%%%%%%%%%%%%%%%%%%%%%%%%%%%%%%%%%%%%%%%%%%%%%%%%%%
\section*{References}
\bibliography{refFourier.bib}
%%%%%%%%%%%%%%%%%%%%%%%%%%%%%%%%%%%%%%%%%%%%%%%%%%%%%%%%
%%%%%%%%%%%%%%%%%%%%%%%%%%%%%%%%%%%%%%%%%%%%%%%%%%%%%%%%

%%%%%%%%%%%%%%%%%%%%%%%%%%%%%%%%%%%%%%%%%%%%%%%%%%%%%%%%
%%%%%%%%%%%%%%%%%%%%%%%%%%%%%%%%%%%%%%%%%%%%%%%%%%%%%%%%
\appendix

%%%%%%%%%%%%%%%%%%%%%%%%%%%%%%%%%%%%%%%%%%%%%%%%%%%%%%%%
%%%%%%%%%%%%%%%%%%%%%%%%%%%%%%%%%%%%%%%%%%%%%%%%%%%%%%%%
\section{Determining an effective $t_0$}
\label{sec:appTimeZero}

\begin{figure}
\centering
\includegraphics[width=\columnwidth]{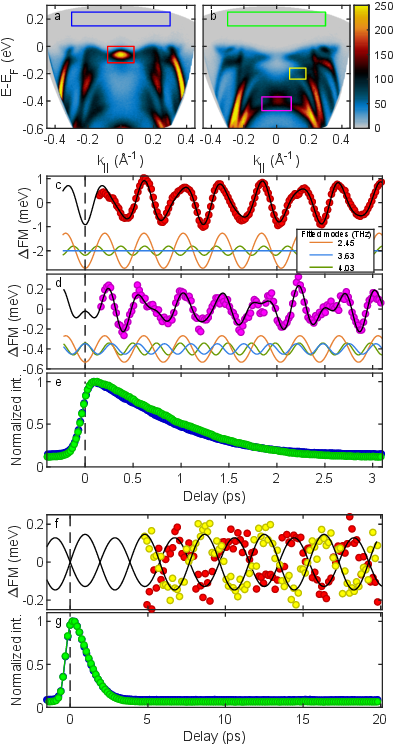}
\caption{(a-b)~Photoemission spectra before $t_0$ for $s$- and $p$-polarized light, respectively. 
Colored boxes indicate the regions of interests for panels (c)-(g).
(c)~FM oscillations in the fast setting for $s$-polarized probe. 
Oscillations are fitted with two modes (black line), as the 3.63~THz mode is absent in this polarization. 
For clarity, individual mode contributions are also presented separately (colored lines, shifted vertically). 
(d)~Same as (c) with $p$-polarized probe and three modes. 
(e)~Normalized population dynamics above $E_F$ for both polarizations.
(f)~FM oscillations in the slow setting for both light polarizations. 
The oscillations are fitted with one mode (black line). 
(g) Normalized population dynamics above $E_F$ for both polarizations.
\tred{The Fourier transforms showing the relevant frequencies are given in Fig.~\ref{fig:determineFreq}c-d}.}
\label{fig:determinet0}
\end{figure}

To determine $t_0$, we choose a region in each spectrum that contains large oscillation amplitudes and adjust the following function to the FM oscillations 
\begin{equation}
    \Delta E(t) = \sum_i A_i \cos\left[ 2 \pi f_i (t - t_0)\right],
\end{equation}
with the amplitude $A_i$ and the frequency $f_i$. 
We keep the frequencies variable for the fast setting and leave the single frequency for the slow setting fixed at the previously determined value. 
The FM oscillations and their fits are presented in Fig.~\ref{fig:determinet0}c,d,f. 
For simplicity, we choose $t_0$ such that 1) the oscillation is a pure sine or cosine function and 2) that it falls within the initial rise of the population dynamics observed far above $E_F$ (Fig ~\ref{fig:determinet0}e,g). 
Conveniently, the three modes observed for the fast setting are well described using a common effective $t_0$, although this is not a requirement for  frequency domain analysis. 
We find cosine oscillations for the high frequency modes for both polarizations (Fig ~\ref{fig:determinet0}c,d), whereas they are sine oscillations for the low frequency mode (Fig ~\ref{fig:determinet0}f).
The sine is in agreement with a previous trARPES work ~\cite{Hein2020}, although other works reported different phases~\cite{Drueke2021,Ji2021a}. 
We emphasize that our methodology is not well suited to determine the phase of the low frequency mode precisely. 
The determination of the phase is challenging in the slow setting due to the difficulty to isolate the oscillations within the first 4~ps and uncertainties of the mode frequency. 
For example, our dataset can also be reasonably described using a cosine function with a slightly different frequency and a $t_0$ that still falls within the intensity rising edge.

%%%%%%%%%%%%%%%%%%%%%%%%%%%%%%%%%%%%%%%%%%%%%%%%%%%%%%%%
%%%%%%%%%%%%%%%%%%%%%%%%%%%%%%%%%%%%%%%%%%%%%%%%%%%%%%%%
\section{Effects of a linearly chirped probe pulse}
\label{sec:appMathChirp}

Following the formalism in Ref.~\cite{Jelovina2018}, a linearly-chirped pulse can be expressed by its time-dependent electric field $E(t)$,
\begin{equation}
    E(t)=\frac{E_0}{(1+\eta^2)^\frac{1}{4}} \exp{\left(\frac{-t^2}{2 \tau^2}\right)} e^{i\Phi(t)}
    \label{eqchirp}
\end{equation}
with
\begin{equation}
    \tau=\Delta t_\text{bare} \sqrt{\frac{1+\eta^2}{4\ln 2}}
\end{equation}
and
\begin{equation}
    \Phi(t)=\omega_0 t +\frac{\eta}{2\tau^2}t^2
\end{equation}
where $\omega_0$ is the central frequency and $\eta$ is the chirp parameter. 
In absence of chirp ($\eta=0$), $E_0$ is the electric field amplitude and $\Delta t_\text{bare}$ is the time resolution, defined as the FWHM of the intensity $|E(t)|^2$. 
This chirp pulse can be expressed in the time-Fourier domain with a Wigner distribution $W(\omega,t)$~\cite{Paye1993}
as
\begin{equation}
\label{eq:Wigner}
W(\omega,t) = E_0^2 \sigma_t \sqrt{\pi} 
e^{\frac{-t^2}{\sigma_t^2}} \cdot
e^{\frac{-(\omega-\omega_0)^2}{\sigma_E^2}} \cdot
e^{2(\omega-\omega_0) \eta t }
\end{equation}
with 
\begin{equation}
 \sigma_t= \frac{\Delta t_\text{bare}}{ \sqrt{4\ln 2}} 
\end{equation}
and
\begin{equation}
\sigma_E=\frac{1}{\sigma_t \sqrt{1+\eta^2} }.
\end{equation}
In the limit of small chirp parameter $\eta$, we can approximate the distribution as
\begin{equation}
W(\omega,t) = W_\text{even}(\omega,t)+W_\text{odd}(\omega,t)
\end{equation}
with
\begin{equation}
W_\text{even}(\omega,t) = E_0^2 \sigma_t \sqrt{\pi} 
e^{\frac{-t^2}{\sigma_t^2}} \cdot
e^{\frac{-(\omega-\omega_0)^2}{\sigma_E^2}}
\end{equation}
and
\begin{equation}
W_\text{odd}(\omega,t) = W_\text{even}(\omega,t) \cdot
2(\omega-\omega_0) \eta t.
\end{equation}
The time-even component is simply the two-dimensional Gaussian expected in the absence of chirp. 
The main contribution from linear chirp is the time-odd component. 

To evaluate the effect of a probe pulse on the measured intensity $I(\omega,t)$, we consider the two-dimensional convolution of the time-dependent spectral intensity $S(\omega,t)$ with this Wigner distribution $W(\omega,t)$, i.e.
\begin{equation}
    I(\omega,t)=\left(S \ast W\right) (\omega,t).
\end{equation}
For simplicity, we assume that the time dependence of $S(\omega,t)$ is described by a single cosine oscillation, i.e. it is written as $S(E,t)=S(E,\cos(\Omega t))$, a time-even function.
The convolution of even functions leads to an even function, while the convolution of odd and even functions leads to an odd function. 
This implies that $\left(S \ast W_\text{even}\right)$ is a time-even function and any oscillations will have a cosine dependence $\cos(\Omega t)$, to first order. 
This term corresponds to the result in the absence of chirp. 
On the other hand,  $\left(S \ast W_\text{odd}\right)$ is a time-odd function and any oscillations will have a sine dependence $\sin(\Omega t)$, to first order. 
This term is a direct consequence of linear chirp.
We therefore conclude that linear chirp leads to artificial oscillations that are $\pm\pi/2$ out-of-phase from the intrinsic oscillations.

%%%%%%%%%%%%%%%%%%%%%%%%%%%%%%%%%%%%%%%%%%%%%%%%%%%%%%%%
%%%%%%%%%%%%%%%%%%%%%%%%%%%%%%%%%%%%%%%%%%%%%%%%%%%%%%%%
\section{Frequency domain representations and FDARPES composite spectra}
\label{sec:appFDARPES}

\begin{figure}[H]
\centering
\includegraphics[width=0.78\columnwidth]{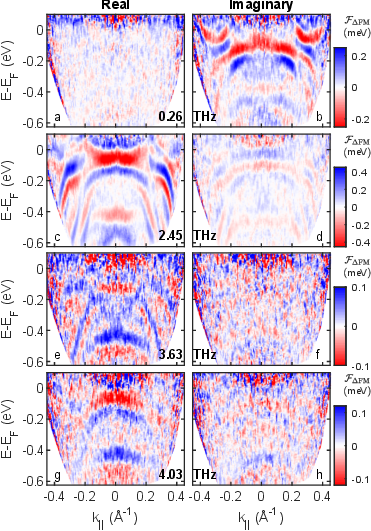}
\caption{FM analysis of all modes in $p$-polarized spectra. 
The real (imaginary) part of the Fourier components is presented in the left (right) column in panels a,c,e,g (b,d,f,h).}
\label{figApp:FmapCOMppol}
\end{figure}

\begin{figure}[H]
\centering
\includegraphics[width=0.78\columnwidth]{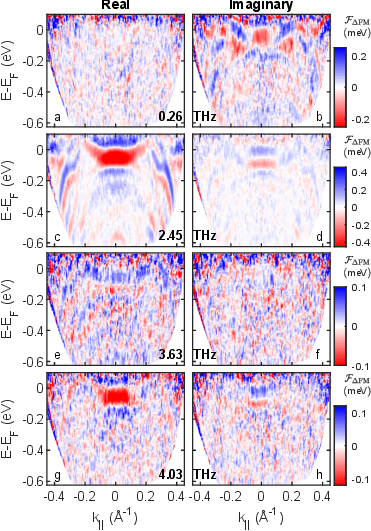}
\caption{FM analysis of all modes in $s$-polarized spectra. 
The real (imaginary) part of the Fourier components is presented in the left (right) column in panels a,c,e,g (b,d,f,h).}
\label{figApp:FmapCOMspol}
\end{figure}

\begin{figure}[H]
\centering
\includegraphics[width=0.78\columnwidth]{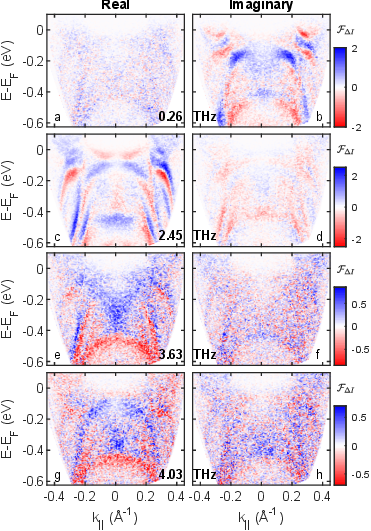}
\caption{PI analysis of all modes in $p$-polarized spectra.  
The real (imaginary) part of the Fourier components is presented in the left (right) column in panels a,c,e,g (b,d,f,h).}
\label{figApp:FmapIntppol}
\end{figure}

\begin{figure}[H]
\centering
\includegraphics[width=0.78\columnwidth]{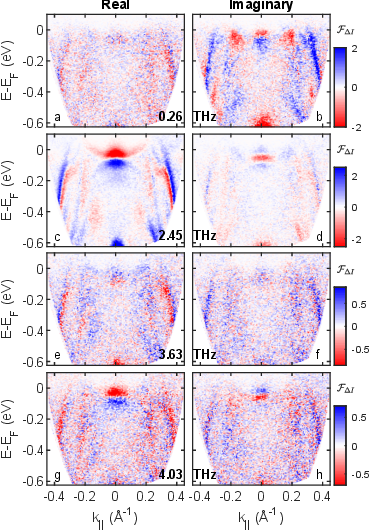}
\caption{PI analysis of all modes in $s$-polarized spectra.  
The real (imaginary) part of the Fourier components is presented in the left (right) column in panels a,c,e,g (b,d,f,h).}
\label{figApp:FmapIntspol}
\end{figure}

\begin{figure}[H]
\centering
\includegraphics[width=\columnwidth]{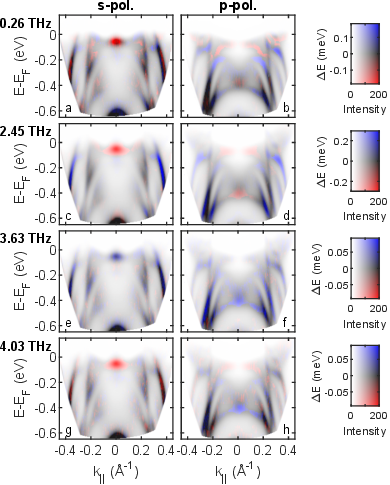}
\caption{Composite FDARPES spectra for all modes using FM analysis. The left (right) columns shows results for $s$ ($p$) polarization in panels a,c,e,g (b,d,f,h).}
\label{figApp:CompositeCOM}
\end{figure}

\begin{figure}[H]
\centering
\includegraphics[width=\columnwidth]{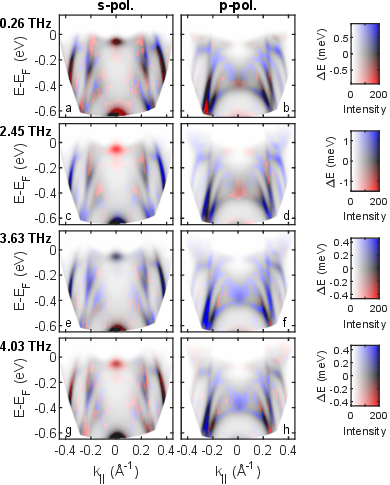}
\caption{Composite FDARPES spectra for all modes using PI analysis. The left (right) columns shows results for $s$ ($p$) polarization in panels a,c,e,g (b,d,f,h).}
\label{figApp:CompositeInt}
\end{figure}

Below we present the full results for all detected modes and $s$- and $p$-polarized probe.
Fig~\ref{figApp:FmapCOMppol} and Fig~\ref{figApp:FmapCOMspol} show the results of the FM analysis.
The results of the PI analysis are shown in Fig~\ref{figApp:FmapIntppol} and Fig~\ref{figApp:FmapIntspol}.
Fig.~\ref{figApp:CompositeCOM} and Fig.~\ref{figApp:CompositeInt} show composite FDARPES spectra derived from FM and PI analysis, respectively. 

\clearpage

\end{document}